\documentclass[aps,pre,twocolumn,floatfix,showpacs]{revtex4}
\usepackage{graphicx}
\usepackage{bm}
\bibstyle{apsrev.bst}

\begin{document}
\title{Stability domains of actin genes and genomic evolution}
\author{E. Carlon}
\affiliation{Interdisciplinary Research Institute, Cit\'e
Scientifique BP 60069, F-59652 Villeneuve d'Ascq, France}
\affiliation{Ecole Polytechnique Universitaire de Lille,
Cit\'e Scientifique, F-59655 Villeneuve d'Ascq, France}
\affiliation{Institute for Theoretical Physics, Katholieke Universiteit
Leuven, Celestijnenlaan 200D, B-3000 Leuven, Belgium}
\author{A. Dkhissi}
\affiliation{Interdisciplinary Research Institute, Cit\'e
Scientifique BP 60069, F-59652 Villeneuve d'Ascq, France}
\author{M. Lejard Malki}
\affiliation{Interdisciplinary Research Institute, Cit\'e
Scientifique BP 60069, F-59652 Villeneuve d'Ascq, France}
\author{R. Blossey}
\affiliation{Interdisciplinary Research Institute, Cit\'e
Scientifique BP 60069, F-59652 Villeneuve d'Ascq, France}
\date{\today}

\begin{abstract}
In eukaryotic genes the protein coding sequence is split into several
fragments, the exons, separated by non-coding DNA stretches, the introns.
Prokaryotes do not have introns in their genome.  We report the
calculations of stability domains of actin genes for various organisms
in the animal, plant and fungi kingdoms. Actin genes have been chosen
because they have been highly conserved during evolution.  In these
genes all introns were removed so as to mimic ancient genes at the time
of the early eukaryotic development, i.e. before introns insertion.
Common stability boundaries are found in evolutionary distant organisms,
which implies that these boundaries date from the early origin of
eukaryotes. In general boundaries correspond with introns positions
of vertebrates and other animals actins, but not much for plants and
fungi.  The sharpest boundary is found in a locus where fungi, algae
and animals have introns in positions separated by one nucleotide only,
which identifies a {\it hot-spot} for insertion. These results suggest
that some introns may have been incorporated into the genomes through a
thermodynamic driven mechanism, in agreement with previous observations
on human genes. They also suggest a different mechanism for introns
insertion in plants and animals.

\end{abstract}


\maketitle

\newcommand{\ul}{\underline}
\newcommand{\bc}{\begin{center}}
\newcommand{\ec}{\end{center}}
\newcommand{\be}{\begin{equation}}
\newcommand{\ee}{\end{equation}}
\newcommand{\ba}{\begin{array}}
\newcommand{\ea}{\end{array}}
\newcommand{\beqn}{\begin{eqnarray}}
\newcommand{\eeqn}{\end{eqnarray}}

\section{Introduction}
\label{sec:intro}

\begin{figure}[b]
\includegraphics[width=8cm]{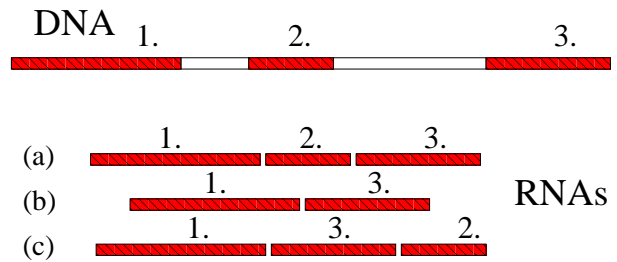}
\caption{Through the mechanism of alternative splicing different RNAs and
thus different proteins can be obtained from the same intron-containing
DNA sequence. Typically, these different RNAs are synthesized in different
tissues.  In the example shown here three different RNAs are formed from
the same DNA sequence: exons are colored, while introns are white. (a)
Exons are assembled following the same order as in the DNA sequence. (b)
Exon 2 is skipped. (c) Exons 2 and 3 are in reversed order compared to
the DNA sequence.}
\label{alternative_splicing}
\end{figure}

Differently from their prokaryotic counterparts, the large majority of
eukaryotic genes are split. The parts of the gene which carry the genetic
code from which the proteins are synthesized, the exons, are interrupted
by long stretches of ``junk DNA", the introns \cite{albe02}.  Much is
still uncertain about introns and in general about junk DNA.  There is
however a clear advantage for a gene of hosting introns: different
mRNAs and henceforth different proteins can be synthesized from the
same gene through a mechanism known as {\it alternative splicing}
(see Fig. \ref{alternative_splicing}). In different tissues of a
multicellular organism the mRNAs are synthesized by placing the exons in
different order or by skipping some of them. This produces quite similar,
but not identical, proteins. Alternative splicing is responsible for the
appearance of slightly different proteins, say, in brain and in liver both
encoded by the same gene.

The origin of introns has triggered quite some debate in the past
years. The discussion was polarized into two different viewpoints: the
``introns early" \cite{gilb78} and the ``introns late" \cite{cava85}
theories. The introns late viewpoint states that introns came ``late" in
the evolution, say after the separation between eukaryotic and prokaryotic
kingdoms. Ancient genomes, like nowadays bacteria, had no introns. During
evolution introns were inserted at some positions in the coding sequence
of eukaryotes. Bacteria did not get introns in order to keep their genome
short. According to the introns early perspective introns were already
present in ancient genomes. In these genomes mini-genes were separated
by junk DNA sequences. Complex genes appeared during evolution when
the mini-genes were assembled together. 

Although the issue is not completely settled yet, there is a widespread
agreement about the fact that most of introns were inserted late in the
genome, except for few which can have a very old origin \cite{roy02}. The
question that remain unanswered is: through which mechanism introns
were inserted into the genes? Did they target some specific stretches
of sequences or their insertion was a random process?

%%%%%%%%%%%%%%%%%%%%%%%%%%%%%%%%% FIG_01 %%%%%%%%%%%%%%%%%%%%%%%%%%%%%%%%%%%
%%%%%%%%%%%%%%%%%%%%%%%%%%%%%%%%% FIG_01 %%%%%%%%%%%%%%%%%%%%%%%%%%%%%%%%%%%

In a previous paper \cite{carl05} we suggested that some introns may
have targeted and got inserted in specific regions of the gene because
of some physical stability properties of these regions. DNA is an
inhomogeneous polymer: sequences richer in CG nucleotides are more stable
than AT rich regions since CG pairs form three hydrogen bonds, while AT
only two.  Using a random set of 80 human genes we found that there is
a strong correlation between intron positions and stability boundaries
\cite{carl05}, to be defined more precisely in the next section.

The aim of this paper is to investigate further on this issue. We consider
here a single gene, the actin, and analyze its stability on animals,
plants and fungi. Although originating from a single gene in a common
ancestor, actin genes have diversified during evolution. Several different
actins are present in the genome of a given eukaryote. By analyzing
the stability properties of genes belonging to a common family we gain
insight on mechanisms of introns insertion. We will show that common
stability boundaries are found in actin sequences of species belonging
to different kingdoms. This implies that the boundaries observed in
this and in previous work \cite{carl05} have a very remote origin,
dating back to the development of early eukaryotes and supports previous
observations that stability boundaries may have influenced the insertion
of at least some introns.  An extensive discussion of the consequences
of our findings is given in the final section of this paper.

\section{Thermodynamic stability}
\label{sec:stability}

When a double helical DNA in solution is brought to a sufficiently high
temperature the two strands dissociate, or melt.  DNA oligonucleotides
of 20-30 base pairs melt at a single temperature.  This temperature can
be estimated using the nearest-neighbor model from which one computes
Gibbs free energies, enthalpies and entropies of melting. Quite some
effort has been dedicated in the past years to an accurate determination
of these thermodynamic parameters (see e.g. \cite{sant98} and references
therein), due to importance of DNA melting and of the reverse transition,
the DNA hybridization, in many biotechnological processes.  The melting
temperature depends, besides on the sequence composition, on salt
concentration and pH of the solution.

If the sequences are sufficiently long, DNA melting becomes a multistep
process \cite{wart85}. Regions of the sequence which are GC-richer
will melt at higher temperatures compared to GC-poorer regions. The
interesting quantities to calculate in this case are multiple partial
melting temperatures and at the same time one needs to determine the
regions of the sequence which melt at those temperatures. In order to
perform such type of calculation various statistical mechanical models
have been developed \cite{wart85,peyr89}.  The calculations presented in
this paper are based on the Meltsim algorithm \cite{marx98}, in which
a DNA configuration is approximated by a sequence of non-interacting
loops and helical segments according to the Poland-Scheraga model
\cite{pola66,pola70}. In this approach each base pair is in two
possible states either open ($\theta_i=0$) or closed ($\theta_i=1$),
where $\theta$ defines the order parameter and $i$ is an index running
over all base pairs of a sequence ($i=1,2 \ldots N$).  In the Meltsim
algorithm recursion relations \cite{pola74} and an approximation for
the closed loops entropy \cite{fixm77} allow a rapid computation of
the opening/closing probability at any given temperature for chains of
several thousands base pairs.

Computations based on Poland-Scheraga
model have been quite popular in the past years
\cite{yera00,yera02,gare03,tost05,metz05,colu07a,ever07}. Yeramian {\it
et al.} analyzed the genomes of {\it S. Cerevisiae} (Yeast) \cite{yera00}
and of {\it P. Falciparum} \cite{yera02} and identified genes on the
basis of thermodynamic signals obtained from the melting analysis. The
effect of mismatches \cite{gare03} and of disorder \cite{colu07a} on
DNA melting have also been discussed.  A recent study has produced the
melting map of the whole human genome \cite{liu07_sh}.

The other popular model for studies of the thermodynamics of
the DNA is the Peyrard-Bishop model \cite{peyr89,peyr06},
which has attracted quite some attention in recent years
\cite{camp98,cocc99,barb03,joy05,mich06,webe06}. This model is probably
more accurate on shorter length scales as a configuration is identified
by the distances between complementary bases and not by a simple boolean
variable ($\theta = 0,1$) as in the Poland-Scheraga picture.  However for
the purposes of calculating stability properties which involve melting
domains of about hundred basis the Poland-Scheraga model is good enough.
Programs like Meltsim have been fine tuned to fit experimental data
\cite{bizz98,marx98}.  Interestingly, the thermodynamic boundaries found
in the Meltsim approach in a previous paper \cite{carl05} have also
been found in an analysis of the Peyrard-Bishop model \cite{joy05}. This
shows that the properties discussed here are robust and model independent.

%%%%%%%%%%%%%%%%%%%%%%%%%%%%%%%%% FIG_02 %%%%%%%%%%%%%%%%%%%%%%%%%%%%%%%%%%%
\begin{figure*}[t]
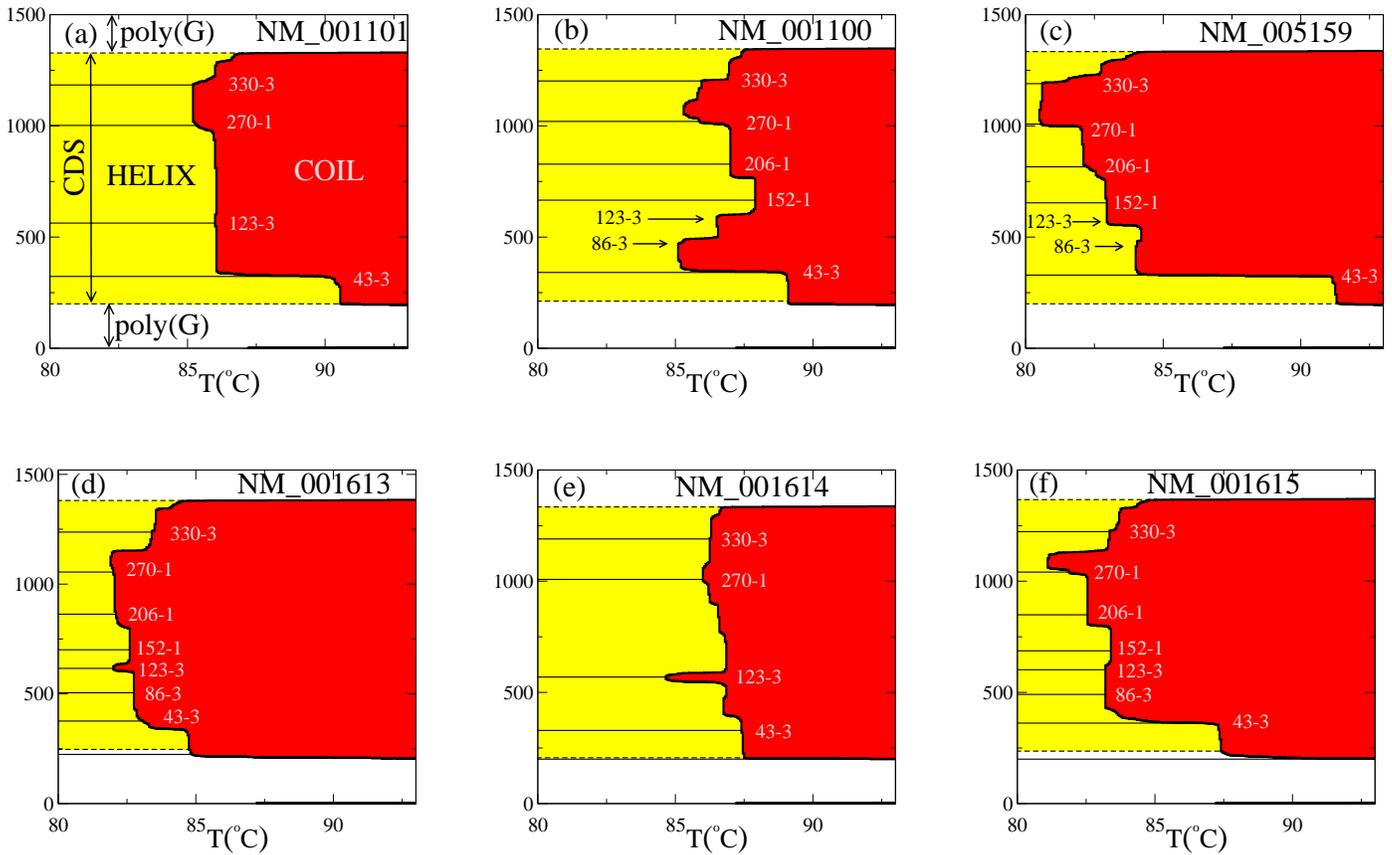

\hspace*{-10mm}
\includegraphics[width=5.4cm]{NM_001101_z.eps}
\ \ \ \ \ \ \ \ \
\includegraphics[width=5.4cm]{NM_001100_z.eps}  
\ \ \ \ \ \ \ \ \
\includegraphics[width=5.4cm]{NM_005159_z.eps}

\vspace*{8mm}
\hspace*{-10mm}
\includegraphics[width=5.4cm]{NM_001613_z.eps}
\ \ \ \ \ \ \ \
\includegraphics[width=5.4cm]{NM_001614_z.eps}
\ \ \ \ \ \ \ \
\includegraphics[width=5.4cm]{NM_001615.eps}
\caption{(Color online) Melting domains for {\it H. Sapiens}
actins. GenBank entries: (a) NM\_001101 (actin $\beta$ ACTB), (b)
NM\_001100 (actin $\alpha_1$, skeletal muscle ACTA1), (c) NM\_005159
(actin $\alpha$, cardiac muscle, ACTC1) (d) NM\_001613 (actin $\alpha_2$,
ACTA2) (e) NM\_001614 (actin $\gamma_1$, ACTG1) (f) NM\_001615 (actin
$\gamma_2$, ACTG2). The sequences have high similarity to those of all
other vertebrates actins, for which almost identical melting patterns
are found.  Hence only human actins are shown as representatives
to those of all the vertebrates. In the plots the temperature is in
the $x$-axis and the sequence position in the $y$-axis. 
The thick solid lines separate the low temperature helix domain 
from the high temperature coiled state.
Horizontal dashed
lines indicate the boundaries of the CDS and the solid lines the intron
positions for the given sequence. Arrows point to the intron positions
found in homologous sequences.}
\label{FIG_ACTBH}
\end{figure*}
%%%%%%%%%%%%%%%%%%%%%%%%%%%%%%%%% FIG_02 %%%%%%%%%%%%%%%%%%%%%%%%%%%%%%%%%%%

In this paper we have used the same set of thermodynamic parameters as
in Ref. \cite{blos03}.  In order to estimate the thermal stability
boundaries, we proceed as follows. Starting from sufficiently
low temperatures in which the whole chain is in an helical state,
we increase the temperature at a constant small step $\Delta T$
($=0.01^\circ$C in the calculation). At each point the configuration
of the chain is calculated and the boundaries between helix and coil
regions recorded. To discriminate between a helical and a coiled region
we calculate the average value of $\theta_i$ at a given temperature and
define the boundary as the point separating a $\theta > 1/2$ domain from
a $\theta < 1/2$ domain.

\begin{figure*}[t]
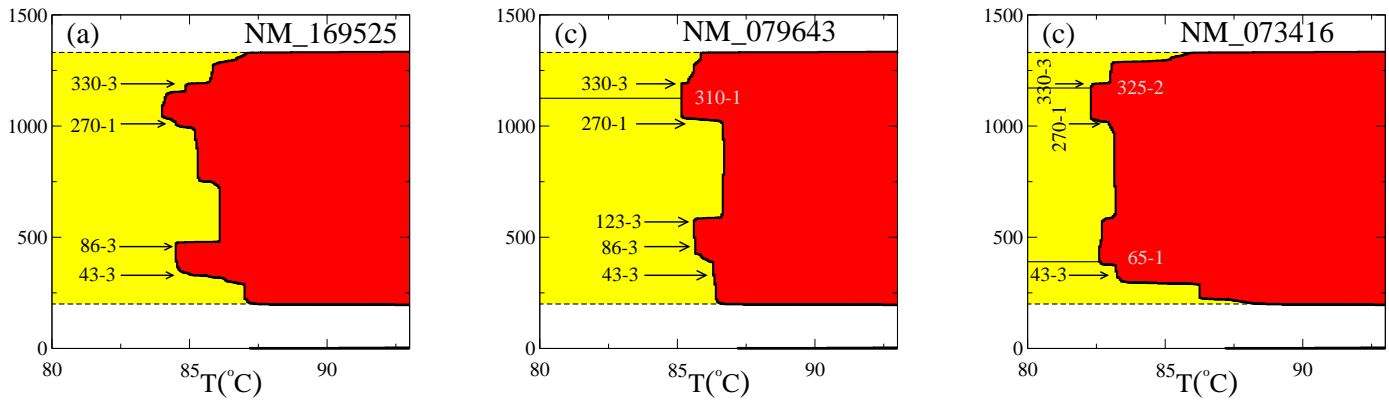

\hspace*{-10mm}
\includegraphics[width=5.4cm]{NM_169525.eps}
\ \ \ \ \ \ \ \ \
\includegraphics[width=5.4cm]{NM_079643.eps}
\ \ \ \ \ \ \ \ \
\includegraphics[width=5.4cm]{NM_073416.eps}
\caption{(Color online) Melting domains for actin genes of {\it D. 
Melanogaster}. 
The area descriptions are the same as in Fig. \ref{FIG_ACTBH}.
(a,b) and {\it C. Elegans} (c). GenBank entries:
(a) NM\_169525 (Act87E), (b) NM\_079643 (Act88F) and (c) 
NM\_073416.}
\label{FIG_ACT_Dros}
\end{figure*}

Typical outputs of the calculations are shown in the graphs of
Fig. \ref{FIG_ACTBH}, \ref{FIG_ACT_Dros}, \ref{FIG_ACT_Arabidopsis}
and \ref{FIG_ACT_Fungi}. In these graphs the x-axis is the temperature,
while the y-axis represents the position along the sequence. For each
gene we considered only the coding sequence (CDS) with all introns
removed. This is the so-called complementary DNA (cDNA), which is a
double stranded copy of the mRNA and can be obtained from it in laboratory
through reverse transcription. For the purposes of inferring information
on genome evolution we can look at cDNA as an old gene before introns
were inserted.  In order to avoid boundary effects, i.e. dissociation
dominated by the opening of forks at the edges, we have enclosed the
sequences by two stretches of poly(G) of two hundred nucleotides each
(a poly(G) sequence is a stretch of DNA composed only of nucleotides
G, in this case the sequence referred to is double-stranded with one
strand containing only G's while the other strand contains only C's).
These stretches have high melting temperature, hence they dissociate
well beyond the melting temperatures of the CDS.  The solid thick lines
in Figs. \ref{FIG_ACTBH}, \ref{FIG_ACT_Dros}, \ref{FIG_ACT_Arabidopsis}
and \ref{FIG_ACT_Fungi} separate the coiled from the helical regions
(to the right and to the left of the curve, respectively).  Due to strong
cooperativity \cite{blak87}, the DNA melts through few sharp transitions
involving the dissociation of hundred of base pairs simultaneously. Hence,
only few stability domains are found in the analysis of a sequence of
1000 base pairs.

\section{Actin}
\label{sec:actin}

Actin proteins play a central role in eukaryotes. Actin filaments
constitute the cytoskeleton of all eukaryotic cells, and are the site of
interactions with many other proteins, as for instance motor proteins or
actin-bundling proteins \cite{albe02}.  A mutation in a specific actin
protein site may result in a change in its interactions with several
proteins that bind near the mutated site. While the mutation can
favor the interaction with one specific protein, it is likely disrupt
interactions with many other proteins. Hence, in order to maintain the
multiple interactions with all its partners, actin proteins have been
highly conserved during evolution. Obviously there is lower conservation
at the gene level compared with the conservation of amino acid sequence for
the corresponding protein, as the genetic code is degenerate and multiple
codons encode for the same amino acid. For actin there is roughly a 80\% of
sequence conservation between human and yeast ({\it S. Cerevisiae}) genes,
while 95\% conservation of amino acids in the proteins \cite{albe02}.

%%%%%%%%%%%%%%%%%%%%%%%%%%%%%%%%%%%%%%%%%%%%%%%%%%%%%%%%%%%%%%%%%%%%%%%%%%%
\begin{table*}[t]
\caption{Table of introns positions in Actin genes for vertebrates and
green plants. The label refers to the codon position following the 
table of Ref. \cite{bhat97}. The three positions surrounded by a box are 
those for which a stability boundary was found.}
\begin{ruledtabular}
\begin{tabular}{c|ccccccccc}
	    & 20-3 & \fbox{43-3}& 86-3& 123-3& 152-1& 206-1& \fbox{270-1}& 
	    \fbox{330-3} & 356-3\\
% \smallskip
\hline
Vertebrates &      &     &     &      &      &      &      &       &\\
$\alpha_1$    &      &  x  &     &      &   x  &   x  &   x  &   x   &\\
$\beta$     &      &  x  &     &  x   &      &      &   x  &   x   &\\
$\gamma_1$    &	   &  x  &     &  x   &      &      &   x  &   x   &\\
$\alpha_2$  &	   &  x  &  x  &  x   &   x  &   x  &   x  &   x   &\\
$\gamma_2$  &	   &  x  &  x  &  x   &   x  &   x  &   x  &   x   &\\
\hline
Green plants & x  &     &     &      &   x  &      &      &       & x 
\end{tabular}
\end{ruledtabular}
\label{table_act_vert}
\end{table*}
%%%%%%%%%%%%%%%%%%%%%%%%%%%%%%%%%%%%%%%%%%%%%%%%%%%%%%%%%%%%%%%%%%%%%%%%%%%%

%%%%%%%%%%%%%%%%%%%%%%%%%%%%%%%%% FIG_03 %%%%%%%%%%%%%%%%%%%%%%%%%%%%%%%%%%%
%%%%%%%%%%%%%%%%%%%%%%%%%%%%%%%%% FIG_03 %%%%%%%%%%%%%%%%%%%%%%%%%%%%%%%%%%%

Genomic evolution is believed to have occurred mainly through gene
duplication and mutations \cite{albe02}: At a given time an error
in the replication of DNA produces two copies of a gene, which are
inherited to a daughter cell. These genes further evolve separately
accumulating different point mutations and thus diverging in time.
As the process is repeated one obtains from a single ancestor gene a
family of closely related genes.  In vertebrates there are three classes
of actins \cite{albe02} known as $\alpha$, $\beta$ and $\gamma$ actins.
The $\alpha$ actins are found in muscle cells, while the $\beta$ and
$\gamma$ are found in non-muscle cells.  Plant actins also form a large
family with even more genes than in vertebrates. For instance more than
10 different actin genes have been identified in the genome of the plant
{\it Arabidopsis Thaliana}.

\subsection{Intron positions}

The Table \ref{table_act_vert} compares the intron positions of
vertebrates and land plants. A more complete table which contains $56$
different intron positions for actins of different organisms can be
found in Ref. \cite{bhat97}.  In total there are $7$ intron positions
for vertebrates actins. These positions are labeled, following the
notation of Ref. \cite{bhat97}, by two numbers. The first number refers
to the codon in the sequence and the second one (between 1 and 3)
indicates where the intron is inserted in the codon. A 3 signifies
that the intron is inserted after the third nucleotide of the codon,
hence the intron does not break the codon.  The codon numbers are given
with respect to a reference sequence, which is the $\alpha$ actin of
vertebrates. Although plants have more actin isoforms than vertebrates,
somewhat quite surprisingly they have only 3 introns positions, one of
which (152-1) is in common with the vertebrates lineage.

\subsection{Melting domains for animal actins}

We start with the description of melting domains in animal
actins. Although Fig. \ref{FIG_ACTBH} shows exclusively human actins,
we found very similar melting profiles also in $\alpha$, $\beta$ and
$\gamma$ actins of other vertebrates as: {\it Canis familiaris} (dog),
{\it Bos Taurus} (cow), {\it Danio Rerio} (zebrafish), {\it Gallus Gallus}
(chicken) etc\ldots Hence, the conclusions drawn from the analysis of
Fig. \ref{FIG_ACTBH} are probably valid for all vertebrates.

Figure \ref{FIG_ACTBH}(a) shows the melting behavior of the
human actin $\beta$ (GenBank entry NM\_001101). This sequence has
four introns at positions 43-3, 123-3, 270-1 and 330-3 (see Table
\ref{table_act_vert}).  These positions are indicated by horizontal
lines in Fig. \ref{FIG_ACTBH}(a).  In this sequence melting is a three
state process. First, at around $85^\circ$C the exon bounded between
introns 270-1 and 330-3 melts. Next the whole CDS sequence melts except
for a short fragment bounded by the intron at 43-3, which then melts
only beyond $90^\circ$C. There is a remarkable correspondence between
the 43-3 position and a sharp stability boundary. Also positions 270-1
and 330-3 show a similar, although weaker, correspondence.  This $\beta$
actin sequence has already been analyzed in Ref. \cite{carl05} (see Fig. 2
in \cite{carl05}). In that analysis the correspondence with the intron
330-3 was missed because different boundary conditions were used. In
this work the CDS is embedded between two poly(G) stretches, so that
melting inside the sequence is always through the formation of loops
bounded between two helical regions. In Ref. \cite{carl05} some parts of
the untranslated regions bounding the CDS were included in the analysis.
As no stable boundary helical regions were included, part of the melting
in Ref. \cite{carl05} occurred through fork openings from the boundaries.
The inclusion of untranslated regions, i.e. of the original genomic
neighborhood in the analysis, is probably not an optimal choice as these
regions are poorly conserved during evolution. As the aim is to look for
signals from ancient genomes it is better to embed the CDS between two
poly(G) stretches. In this way all the sequences analyzed are treated
on equal footing.

%%%%%%%%%%%%%%%%%%%%%%%%%%%%%%%%% FIG_04 %%%%%%%%%%%%%%%%%%%%%%%%%%%%%%%%%%%
\begin{figure*}[t]
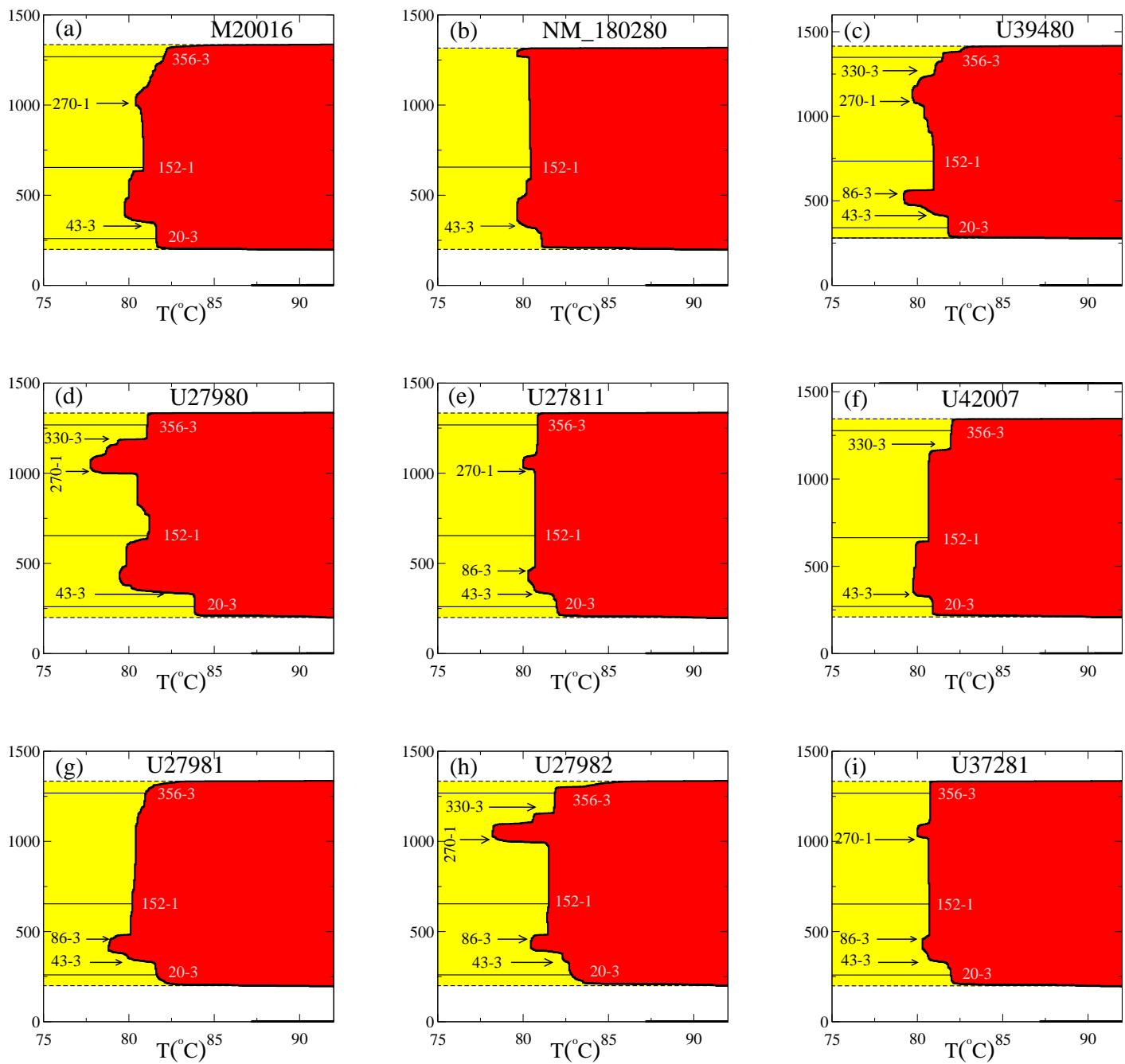

\hspace*{-10mm}
\includegraphics[width=5.5cm]{M20016.eps}
\ \ \ \ \ \ \ \ \
\includegraphics[width=5.5cm]{NM_180280.eps}
\ \ \ \ \ \ \ \ \
\includegraphics[width=5.5cm]{U39480.eps}
 
\vspace*{8mm}
\hspace*{-10mm}
\includegraphics[width=5.5cm]{U27980.eps}
\ \ \ \ \ \ \ \ \
\includegraphics[width=5.5cm]{U27811.eps}
\ \ \ \ \ \ \ \ \
\includegraphics[width=5.5cm]{U42007.eps}
 
\vspace*{8mm}
\hspace*{-10mm}
\includegraphics[width=5.5cm]{U27981.eps}
\ \ \ \ \ \ \ \ \
\includegraphics[width=5.5cm]{U27982.eps}
\ \ \ \ \ \ \ \ \
\includegraphics[width=5.5cm]{U37281.eps}
\caption{(Color online) Melting domains for actin sequences of the {\it
A. Thaliana}. 
The area descriptions are the same as in Fig. \ref{FIG_ACTBH}.
GenBank entries: (a) M20016 (AAc1), (b) NM\_180280 (ACT2),
(c) U39480 (ACT3), (d) U27980 (ACT4,) (e) U27811 (ACT7,) (f) U42007
(ACT8), (g) U27981 (ACT11), (h) U27982 (ACT12) and (i) U37281 (actin 2).}
\label{FIG_ACT_Arabidopsis}
\end{figure*}
%%%%%%%%%%%%%%%%%%%%%%%%%%%%%%%%% FIG_04 %%%%%%%%%%%%%%%%%%%%%%%%%%%%%%%%%%%

The Figs. \ref{FIG_ACTBH}(b-f) show other melting domains for homologous
vertebrates actin genes. These sequences have 4 (e), 5 (b,c) and 7 (d,f)
intron positions. There are three introns positions which are in common
in the vertebrates actins: 43-3, 270-1 and 330-3.  These are also those
for which a correspondence with thermal boundary was found in the $\beta$
gene of Fig. \ref{FIG_ACTBH} (a). The correspondence between thermodynamic
boundaries with the introns positions 43-3 and 270-1 is also observed in
the three other sequences of Fig. \ref{FIG_ACTBH} (b,c,f). Note the very
sharp signal from the 43-3 intron in the case (c). The intron at position
330-3 shows a good correspondence with stability boundaries for the
sequences in Fig. \ref{FIG_ACTBH} (b,c). A much weaker, but noticeable,
correspondence is with the intron at position 206-1 in sequences (c)
and (d). The stability boundary is slightly shifted from the 206-1 in
sequences (d) and (f). The correspondence between intron positions and
thermodynamic boundaries is absent in the sequence (d).

Another interesting feature of vertebrates actins can be seen in
Fig. \ref{FIG_ACTBH}(b). This sequence is the actin $\alpha_1$ which hosts
5 introns in its coding region. These are marked by horizontal lines.
The two remaining of the total 7 intron positions of vertebrates actins,
the 86-3 and 123-3, are indicated by horizontal lines. As it can be seen
from Fig. \ref{FIG_ACTBH}(b), these two positions correspond to stability
boundaries. The correspondence of a stability boundary with the 123-3
is also visible in (c) and (e). In the latter example the 123-3 is a
nucleation site for a small loop.  The sequences of Fig. \ref{FIG_ACTBH}
(d), (e) and (f) show a much weaker correspondence between intron
positions and stability boundaries.

Fig. \ref{FIG_ACT_Dros} shows the melting curves for {\it Drosophila
Melanogaster} (a,b), the fruit fly, and {\it Caenorhabditis Elegans} (c),
a worm. The Drosophila actins have at most one intron in the coding region
either in 15-1 or 310-1. These positions differ from the vertebrates
positions discussed so far.  The sequence shown in Fig. \ref{FIG_ACT_Dros}
(a) has no introns.  The melting analysis however reveals few stability
boundaries close to the positions 43-3, 86-3, 270-1 and 330-3, which
are the introns position of vertebrates actins. The 43-3 and 86-3 are
particularly sharp.  The next Drosophila sequence (b) with one intron at
position 310-1 show a stability boundary close to 270-1 and a weaker one
close to 330-3. Compared to the case in (a), in this sequence the signals
from 43-3 and 86-3 have been lost. However a sharp boundary has appeared
close to the vertebrates intron 123-3.  The {\it C. Elegans} sequence
of Fig. \ref{FIG_ACT_Dros}(c) has two introns at ``new" positions 65-1
and 325-2. As in the previous examples one observes boundaries close to
the 43-3, 123-3, 270-1 and 330-3 positions.

\subsection{Melting domains for plant actins}

Figure \ref{FIG_ACT_Arabidopsis} shows the melting domains for actin genes
of the green plant {\it Arabidopsis Thaliana}. The introns positions
of actins sequences of higher plants are highly conserved (see Table
\ref{table_act_vert}), which indicates that these introns date back to the
early evolution of land plants.  In 3 out of the 9 Arabidopsis sequences
shown Fig. \ref{FIG_ACT_Arabidopsis} (a, d, f) we find a correspondence
of a thermal boundary and an intron at 152-1. This is the intron which is
in common with vertebrates (see Table \ref{table_act_vert}).  As in the
Drosophila and C. Elegans sequences (Fig. \ref{FIG_ACT_Dros}) in general
stability boundaries tend to be found at vertebrates positions 43-3,
86-3, 270-1 and 330-3. In few cases the correspondence is very striking,
as in Fig. \ref{FIG_ACT_Arabidopsis} (d).

\begin{figure}[t]
\includegraphics[width=8.5cm]{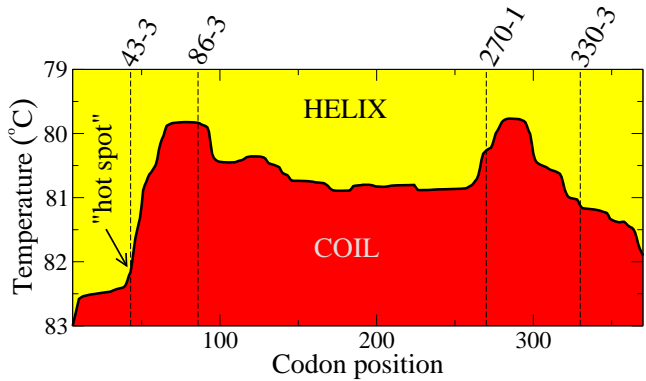}
\caption{(Color online) Average melting curves from 21 green plant
actin genes.  The axes of the diagram are swapped compared to those in
Figs. \ref{FIG_ACTBH} -\ref{FIG_ACT_Arabidopsis}. The horizontal axis
is given in codon position; only the coding sequence is shown. The four
vertical lines denote the major four introns positions found to correlate
with stability boundaries in vertebrates.}
\label{average}
\end{figure}

In order to corroborate these findings we extended the analysis to other
plants. We considered 12 additional actins from {\it Nicotiana Tabacum}
(tobacco, GenBank X63603), {\it Oryza Sativa} (rice, GenBank X15862,
X15863, X15864, X15865), {\it Glycine Max} (soyabean, GenBank J01298,
V00450), {\it Solanum Tuberosum} (potato, GenBank X55749, X55750, X55751,
X55752) and {\it Striga Asiatica} (GenBank U68461, U68462).  With the 9
sequences from {\it A. Thaliana} we have in total 21 plant actin genes.
For each sequence the melting curves were calculated and then averaged.
The result is shown in Fig. \ref{average}. In the graph the $x$ and $y$
axes are reversed compared to the Figures \ref{FIG_ACT_Arabidopsis}.
The $x$-axis is the codon position, and the temperature, now in the $y$
axis, is ordered as increasing from top to bottom.  As reference four of
the most commonly found introns positions for actin genes are indicated
as vertical lines.  The averaging introduces some smoothening, but it 
confirms the existence of a sharp stability boundary close to the 
43-3 position. Two weaker boundaries are found close to the positions 86-3
and 270-1.

%%%%%%%%%%%%%%%%%%%%%%%%%%%%%%%%% FIG_05 %%%%%%%%%%%%%%%%%%%%%%%%%%%%%%%%%%%
%%%%%%%%%%%%%%%%%%%%%%%%%%%%%%%%% FIG_05 %%%%%%%%%%%%%%%%%%%%%%%%%%%%%%%%%%%

\subsection{Melting domains for fungi actins}

To conclude the analysis of the stability behavior of actin genes we
consider now fungi. Fig. \ref{FIG_ACT_Fungi} shows the melting curves
for the budding yeast {\it Saccharomyces Cerevisiae} (a), for {\it
Neurospora Crassa} (b) and for {\it Candida Albicans} (c).  In general
the number of introns and their positions are highly variable in fungi
actin genes: their number vary from 0 to 7 and the positions are most
likely concentrated in the region before the 50th codon.  This can be
seen also in the sequences of Fig. \ref{FIG_ACT_Fungi} two of which have
one intron and one has four. All introns are found before the codon 45.
The melting behavior shown in Fig. \ref{FIG_ACT_Fungi} resembles that
of the previous cases. A sharp stability boundary close to the 43-3
position and some weaker ones appearing close to positions 270-1 and
330-3 in the case (b). This correlation is absent in the case (a).

\subsection{A ``hotspot" for introns insertion}

\begin{figure*}[t]
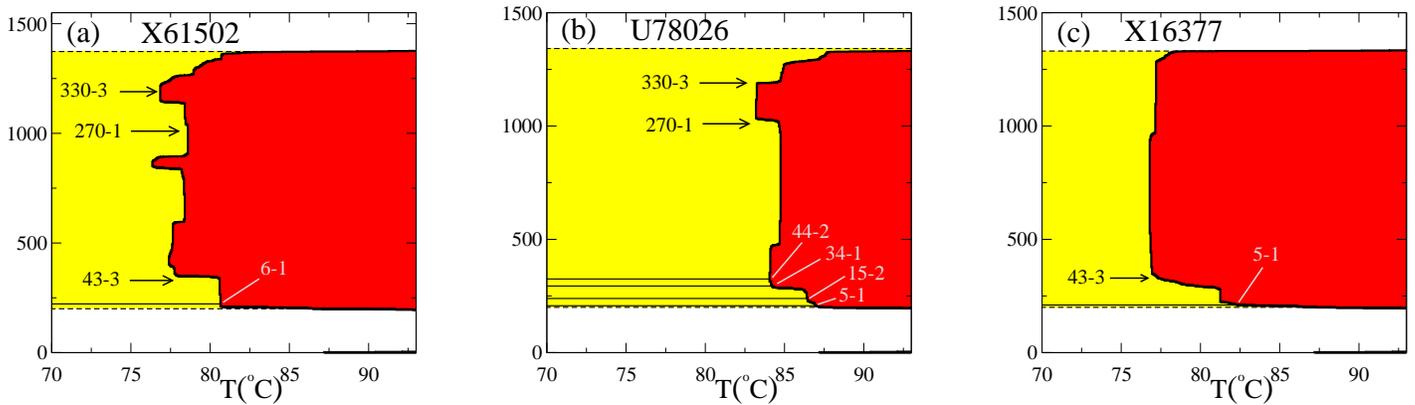

\hspace*{-10mm}
\includegraphics[width=5.5cm]{X61502.eps}
\ \ \ \ \ \ \ \ \
\includegraphics[width=5.5cm]{U78026.eps}
\ \ \ \ \ \ \ \ \
\includegraphics[width=5.5cm]{X16377.eps}
\caption{(Color online) Melting domains for actin sequences of Fungi.
The area descriptions are the same as in Fig. \ref{FIG_ACTBH}.
(a) {\it
Saccharomyces Cerevisiae} (yeast) (b) {\it Neurospora Crassa} and (c)
{\it Candida Albicans}.  GenBank entries: (a) X61502 (Act2), (b) U78026
and (c) X16377 (act1).}
\label{FIG_ACT_Fungi}
\end{figure*}

In Ref. \cite{bhat97} the full table of introns positions in actin genes
gives 56 different positions. There is an interesting remark concerning
the position 43-3, which is shared by all vertebrates actins.  Position
43-3 is relatively common in animals, but it is also the sole intron in
the actin gene of the red alga {\it Chondrus crispus} \cite{bhat97}.
Shifted of a single nucleotide at position 44-1 there is an intron
in the alga {\it Cyanophora Paradoxa} \cite{bhat97}. An intron at
position 44-2 is present in the single-copy actin genes of the fungi
{\it Thermomyces lanuginosa}, {\it Aspergillus Niger}, {\it Neurospora
Crassa} (see Fig.~\ref{FIG_ACT_Fungi}(b)) and {\it Trichoderma Reesei}
\cite{bhat97}. The only other positions with introns separated by a
single nucleotide are at 34-1 and 34-2, which however are only found
in some Fungi.  Hence the 43-3 is a unique site in the actin genes,
which we can refer to as an ``hotspot" for introns insertion.

%%%%%%%%%%%%%%%%%%%%%%%%%%%%%%%%% FIG_06 %%%%%%%%%%%%%%%%%%%%%%%%%%%%%%%%%%%
%%%%%%%%%%%%%%%%%%%%%%%%%%%%%%%%% FIG_06 %%%%%%%%%%%%%%%%%%%%%%%%%%%%%%%%%%%

\section{Discussion}
\label{sec:discussion}

\begin{figure*}[t]
\includegraphics[width=15.5cm]{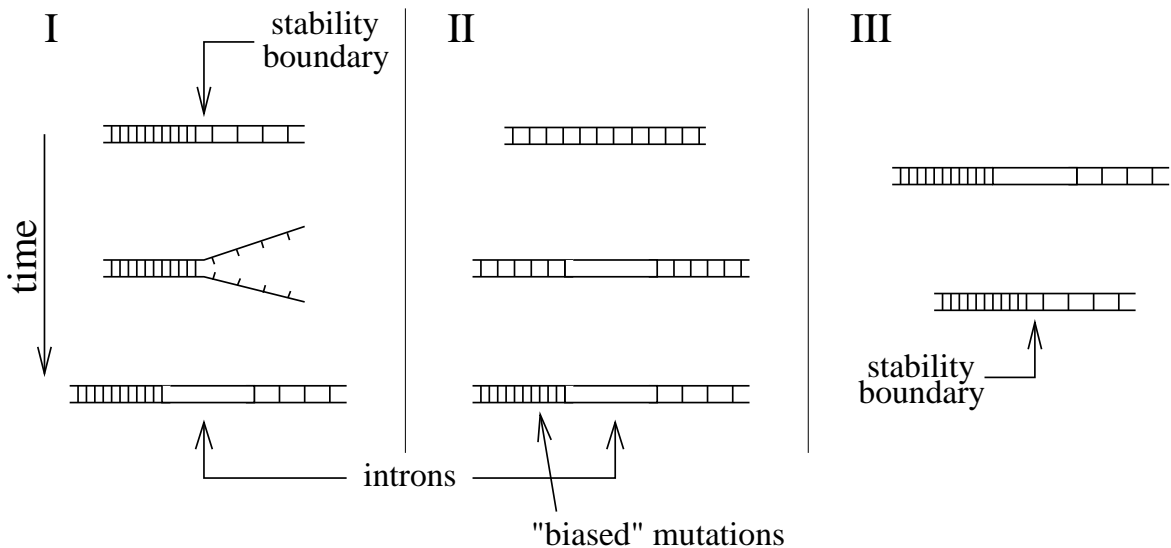}
\caption{Possible scenarios of gene evolutions. (I), (II) introns late
and (III) introns early perspectives. The scheme (I) is the insertion
driven by thermodynamics, supported by the analysis of stability regions
reported in this paper.  The scheme (II) suggests the appearance of a
thermal boundary after intron insertion, a scenario not supported by
the results presented in this paper. Finally the scheme (III) takes
into account the introns early perspective in which the ancient exons
had already different stability properties.}
\label{final_scheme}
\end{figure*}

DNA sequences, which are hundred of base pairs long, tend to melt
through a series of separate temperature steps. Each step consists of
the melting of a region of few hundred of base pairs.  By following the
melting process over a wide temperature interval one can thus identify
separate melting domains, i.e. parts of the sequence which dissociate
at different temperatures. The domain boundaries are points in which
the sequence tend to form in a relatively wide temperature interval a
stable Y-conformation separating a double helix from a coiled region
(see Fig. \ref{final_scheme}).

A previous study \cite{carl05} of about 80 human genes from which introns
are removed and exons linked together revealed that stability domain
boundaries tend to be localized at the end of exons. This correspondence
was found for about 35\% of the exons analyzed.  The correlation was found
to be stronger for a class of so-called housekeeping genes, i.e. those
genes involved in the basic cellular processes. These genes are expressed
in all tissues and have been more conserved during evolution. Actin is
in fact an example of a housekeeping gene.  If one accepts an ``introns
late" viewpoint, the correlation between intron positions and stability
boundaries suggests that some introns were inserted into genes at the ends
of the melting domains in a process driven by thermodynamics.  Such a
process is illustrated in Fig. \ref{final_scheme}(I): an intronless
fragment of a gene has naturally parts which are richer and poorer of
CG nucleotides. When the two strands partially separate they may form
a Y configuration the end of a less-stable domain and the beginning of
a more stable one. Introns may have targeted these fork locations.

Another possibility that may have explained the correlation
between thermodynamic boundaries and introns positions observed in
Ref.~\cite{carl05} is schematically shown in Fig. \ref{final_scheme}(II).
Originally the insertion site does not possess a thermodynamic boundary,
so the intron is inserted through a process which does not depend on
thermodynamics. Once the insertion has taken place and the two exons are
separated by an intron stretch. Mutations may have biased the CG content
on the two exons so that their thermodynamic boundary originated after
the intron insertion. However, this scheme is at odds with the results
presented in this paper. We have indeed shown that boundaries in conserved
positions are found in actin family genes where no introns are present
close to those positions. For instance, in many actins of plants, fungi
and animals there is a sharp stability boundary at the position 43-3
in sequences which have no intron at that position. Hence being found
in plants, animals and fungi sequences, the stability boundary at 43-3
is rather a property of an intronless ancestor actin gene. The same is
true for stability boundaries found in other intronless positions as
for instance the 86-3 and 270-1.

%%%%%%%%%%%%%%%%%%%%%%%%%%%%%%%%% FIG_07 %%%%%%%%%%%%%%%%%%%%%%%%%%%%%%%%%%%
%%%%%%%%%%%%%%%%%%%%%%%%%%%%%%%%% FIG_07 %%%%%%%%%%%%%%%%%%%%%%%%%%%%%%%%%%%

We further speculate on the ``introns early" perspective, i.e. the
possibility that introns were already present in early genomes and were
selectively lost by some species. Our findings then imply that these
introns would have separated early exons (or mini-genes as they are
also referred to) with different stability properties, as shown in the
scheme of Fig. \ref{final_scheme}(III). Although possible, this scenario
seems to be in contradiction with most of the recent phylogenetics based
studies which favor an introns late theory.

In conclusion, our work supports a mechanism given in
Fig. \ref{final_scheme}(I), i.e. a thermodynamic driven introns insertion.
This does not necessarily mean that the actual insertion process took
place through an equilibrium transition with a temperature rise to
$80^\circ$C. First of all the melting temperature depends also on
other salt concentration and pH of the environment.  Moreover, the
boundaries found in the melting analysis, should manifest themselves
also under nonequilibrium conditions. Y-configurations as those
shown in Fig. \ref{final_scheme} can also be generated by mechanical
unzipping of DNA \cite{week05}.
Quite remarkable is the fact that the sharpest boundary in actin genes
(43-3) is also the locus in which introns insertion has been the most
active in evolutionary distant organisms. As we have pointed out introns
have also been found at positions 44-1 and 44-2. This fact is in agreement
with an idea of an insertion driven by thermodynamics. As the boundary
is particularly sharp the mechanism of Fig. \ref{final_scheme}(I) can
have occurred independently on three different families of actin genes.

The correlation between stability boundaries and introns position in
particularly sharp in several sequences analyzed, but in few cases
is absent. We believe that this is due to mutations having erased the
correlation from the ancestral gene sequence. Although actin is highly
conserved as a protein, there is no selective pressure against synonymous
mutations which do not modify the aminiacid sequence. Such mutations
are known to have occurred at a roughly constant rate in all genes of a
given organisms \cite{albe02}. Two genes of the same family in the same
organism have evolved separately and mutations may have accumulated at
higher/lower rates in different parts of the sequence: in some genes the
mutations may have erased the ancient stability boundaries. 

The problem of introns evolution has been widely debated in the
biological literature (for a recent review of the state of the art see
Ref.~\cite{roy06}). Even within the introns late perspective there is no
general consensus on the mechanism of insertion and several possibilites
have been analyzed. For instance Ref.~\cite{roy06} reports 5 different
models of introns insertion. Most of these models in general discuss
the mechanism of insertion without suggesting in which position of
the sequence the insertion would have occurred. One exception is the
protosplice site model \cite{dibb89} which suggests a bias towards a
specific insertion sequence (C/A)AG(G/A) (here C/A denotes a site which
can posses either a nucleotide C or A), referred to as the protosplice
sites. This insertion would have lead to a structure (C/A)AG-intron-(G/A).
The protosplice model and other models for introns insertions are only
partially supported by the analysis of genomic data \cite{roy06}.

Unfortunately, the genomes nowadays investigated have been heavily
reshaped by hundreds of millions of years of evolution and are quite
different from genomes of early eucaryotes.  Hence the answer to the
question of introns origin is not an easy one. Indeed although introns
were discovered 30 years ago, there still an open debate on this issue.
Moreover, evolution may have taken place through complex and diversified
pathways so it is not unlikely that different mechanisms of insertion
have coexisted. Certainly the possibility that also the physical and
thermodynamical stability of the double helix has played a role offers
new insights and stimulates further research in this field.

% \bibliography{/home/staff/enrico/TEX/biblio.bib}
% \bibliography{/home/enrico/TEX/biblio.bib}

\newpage

\end{document}